# Superconductivity at the metal-insulator phase boundary in a bulk nickelate at ambient pressure

Hyo-Bin Ahn[1,*], Xinglong Chen[1,*,#], Hong Zheng[1], Yu Zhang[1], Ramakanta Chapai[1,♠], Yu Li[1], Arashdeep S. Thind[2], Robert F. Klie[2], Michael R. Norman[1], Ulrich Welp[1], J. F. Mitchell[1], and Daniel Phelan[1]

[1]Material Science Division, Argonne National Laboratory, Lemont, Illinois 60439, USA

[2]Department of Physics, University of Illinois Chicago, Chicago, IL 60607, USA

[*] These authors contributed equally to this work

[#] Current affiliation: School of Physics, Southeast University, Nanjing 211189, China

♠ Current affiliation: Department of Physics, Norfolk State University, Norfolk, VA 23504, USA

**Abstract**

The discovery of superconductivity in nickelates has seeded a new field for exploring unconventional superconductivity in transition metal oxides. However, to date superconductivity has only been realized in thin films or under extreme pressure in the bulk. Here we report signatures of superconductivity at *ambient* pressure in *bulk* nickelate single crystals of $(La_{1-x}Pr_x)_4Ni_3O_8$ and $(La_{1-x}Y_x)_4Ni_3O_8$, whose crystal structure comprises interleaving trilayers of square-planar nickel oxide and fluorite-like spacer layers. The parent compound $La_4Ni_3O_8$ exhibits an insulating ground state, where electrons order into intertwined, insulating charge/spin stripes. Substitution of smaller lanthanide ions disrupts this order, eventually leading to a metallic ground-state. We find that superconductivity emerges in a narrow window proximate to the insulator-metal phase boundary, where both metallic and charge-stripe phases co-exist. The observed low volume-fraction superconductivity is non-percolative, suggesting the prospect of filamentary superconductivity nucleated at the boundary between these phases. However, intergrowth defects that approximate the known infinite layer nickelates are observed in TEM, leaving open the possibility that superconductivity resides here rather than in the trilayer matrix. Remarkably, the electronic phase diagrams of both the Y and Pr series coincide when parameterized by the volume of the fluorite-like spacer layers, revealing that this steric parameter profoundly modifies the nickel oxide trilayer electronic structure. Our results identify better understanding of the co-existence region between metallic and charge-ordered phases as a priority for expanding the range of bulk, ambient-pressure nickelate superconductivity and establish spacer-layer engineering as a design tool for exploring this regime of phase competition.

## Introduction

Rare earth nickel oxides (nickelates) have long been targeted as unconventional superconductors since $Ni^{1+}$ is notionally analogous to $S$=1/2 $Cu^{2+}$ - a crucial ingredient of the cuprates[1]. In recent years, two classes of nickelate superconductors have been discovered. The first class, typified by "infinite-layer" 1-1-2 compounds, seem to follow the "cuprate paradigm" insomuch as they contain strongly orbitally polarized $Ni^{1+\delta}$ ions, which have a $3d^{9-\delta}$ electronic configuration like $Cu^{2+\delta}$, in square-planar coordination[2–4]. Although these materials exist in the bulk and as epitaxial thin films, superconductivity has only been reported in thin films[2,4–6], implicating an essential role of strain. A second, perhaps more surprising class has also been reported with $Ni^{3-\delta}$ ($3d^{8-\delta}$) in octahedral coordination in bilayer and trilayer Ruddlesden-Popper phases, seemingly breaking from the cuprate paradigm[7–17]. Here, superconductivity occurs in the bulk only when compressed to extreme pressures (>10 GPa) in diamond anvil cells or again as epitaxial thin films under compressive strain[18,19], except for one report of trace levels of superconductivity in the bilayer Ruddelesden-Popper phase at ambient pressure[20]. Thus, the superconductivity in both classes is extremely sensitive to the arrangement of atoms in the crystalline structure[21], particularly the Ni-O bonds and their symmetry. However, the necessity of thin film epitaxy or diamond anvil cells limits the experiments that can be carried out to fully probe the materials science and underlying physics.

A recurring theme in unconventional superconductivity is that it frequently emerges when forms of charge and/or magnetic ordering - such as charge stripes and spin stripes, or related charge density waves (CDWs) and spin density waves (SDWs) - are suppressed by a tuning parameter that favors charge delocalization: pressure, carrier concentration, etc. This paradigm is observed in Kagome metals[22], cuprates[23], transition metal dicalchogenidesd[24], moire systems[25], organic conductors[26], and heavy fermion materials[27]. As the charge/spin ordering temperature is pushed towards a quantum critical point at $T$=0, a superconducting dome can emerge as a function of the tuning parameter (see Fig. 1(a)). Indeed, experimental reports indicate that the two classes of nickelate superconductors mentioned above exhibit CDW/ SDW (stripe) order nearby their superconducting phases[28–32]. Density wave physics has been argued to be the critical determinant for the Ruddlesden-Popper compounds that superconduct under pressure[33,34]. This suggests that the suppression of charge and spin order instabilities represent a pathway towards design or discovery of additional nickelate superconductors. Here, we demonstrate that non-percolative superconducting regions form in ***bulk*** single crystals of the reduced trilayer nickelate $(La_{1-x}R_x)_4Ni_3O_8$ (R≡Pr or Y, which we refer to as LPNO and LYNO, respectively) under ambient pressure. Notably, superconductivity is found only, for example, in an extremely narrow composition region at the phase boundary where a charge and spin-stripe ordered state is suppressed towards $T$=0 by chemical pressure. Although the observation that these non-percolative superconducting regions are only detected at the phase boundary implicates some role for phase competition, their low volume fraction also raises questions about the potential role of defects, strain, and/or interfaces present in the crystals.

The parent compound of our study, $La_4Ni_3O_8$, has a tetragonal *I4/mmm* crystal structure that consists of trilayers of square-planar Ni-O sheets, as shown in Fig. 1(b) (we refer to this as a 4-3-8 phase); essentially, it can be thought of as a trilayer version of the $LaNiO_2$ (1-1-2) structure, with each trilayer separated by a rare earth fluorite-like spacer layer. Neighboring trilayers are related by the *I*-centering translation. Single crystals of $La_4Ni_3O_8$ possess slightly semiconducting behavior at room temperature, but $La_4Ni_3O_8$ undergoes a 1$^{st}$-order transition into an insulating state at $T_{DW} \approx 105$ K in which it develops static charge/spin order that can be thought of in simple terms as three-period stripes where alternating rows of 1+,1+,2+,1+,1+,2+… nickel ions are oriented diagonal to the Ni-O bond direction[35]. Single crystals of $Pr_4Ni_3O_8$, on the other hand, exhibit metallic behavior. Both La and Pr 4-3-8 have a highly orbitally polarized state with holes in the $d_{x^2-y^2}$ orbitals, reminiscent of the cuprates[36], and their magnetic excitation spectra observed by resonant inelastic X-ray scattering (RIXS) are nearly identical[37]. It has been pointed out that the average nickel valence of 4/3+ yields a *d*-electron count corresponding formally to the over-doped regime of cuprates, which was conjectured to explain why they don't superconduct[36]. Epitaxial thin films with homologous structures consisting of multilayers of four to eight layers have been shown to superconduct[4,38], presumably because the hole concentration, which goes as $1/n$ where $n$ is the number of layers in a multilayer, decreases as $n$ increases, pushing the system into the superconducting regime.

We have previously characterized single crystals of the mixed LPNO system[39] (which was also reported in polycrystalline form[40]), and shown that, instead of forming a clean quantum critical point at the insulator-metal boundary, the transition remains strongly hysteretic and 1st-order until the stripe order is completely suppressed. As charge/spin stripes form near the phase boundary, the system exhibits phase segregation, and the charge/spin-ordered insulator-metal transition can be explained as percolative[39]. Here, we report results from more detailed magneto-transport and susceptibility measurements characteristic of numerous crystals near the insulator-metal transition in LPNO as well as a newly synthesized system, LYNO, for which $Y^{3+}$ substitution is found to be site-selective to the fluorite-like spacer layers in between the trilayers. Superconducting signatures, including a field-dependent resistive transition and a concomitant onset of a diamagnetic contribution to susceptibility, are observed in both series proximate to the insulator-metal transition region. In both systems, the temperature-dependent resistivity of superconducting samples is similar, with the superconducting transition emerging at a temperature below that where the hysteresis associated with the 1$^{st}$-order insulator-metal transition closes, suggesting a common origin of superconductivity despite their different critical temperatures of 7 K and 15 K, respectively. The concomitant diamagnetic response appearing at the same temperatures as the abrupt resistivity downturn, together with the suppression of both features under applied magnetic fields, further supports the presence of superconductivity in these systems. Despite the small superconducting volume fraction (a few percent), as evidenced by the finite residual resistance and weak diamagnetic response, well-defined critical currents are observed in both systems, indicating that non-percolative superconducting regions are present in the crystals. By comparing both Pr and Y systems, it is revealed that they have a common phase diagram as a function of the volume of their fluorite-like spacer layers, offering a new roadmap for exploiting steric effects to impact the electronic behavior of the trilayer electronic subsystem.

**Results**

Figs. 1(c), 1(d), and Extended Data Fig. 1 show the evolution of the temperature-dependent normalized resistance with $x$ of both LPNO and LYNO. It is immediately apparent that the stripe transition at $T_{DW}$ is suppressed in both series with $x$, but at a much faster rate in LYNO, tracking the significantly smaller effective ionic radius of $Y^{3+}$ (1.019 Å) compared to $Pr^{3+}$ (1.126 Å)[41]. Both materials possess compositions with a finite $T_{DW}$ but for which the resistivity appears re-entrant; by that we explicitly mean that the resistivity increases as $T$ is lowered below $T_{DW}$, reaches a maximum, and then drops upon further decreasing temperature. This is likely a consequence of inhomogeneous transport where metallic and insulating regions co-exist with differing temperature-dependent resistivities that lead to a complex temperature dependence rather than a true re-entrant behavior. Significant hysteresis between cooling and warming curves indicates the 1st-order nature of the transition as does the heat capacity[39]. Nevertheless, in both series, the hysteresis loop closes upon further cooling. Abrupt drops were observed in $\mu_0 H$=0 T at $T_C \approx 7$ K [Figs. 2(a) and 2(b)] and $T_C \approx 15$ K [Extended Data Figs. 2(a) and 2(b)) for LPNO ($x$=0.55) and LYNO ($x$=0.11), respectively]. Below we show evidence in magneto-transport, *I-V* characteristics, and magnetic susceptibility that these abrupt resistance drops are signatures of superconductivity.

The temperature dependence of the ***ab***-plane resistivity for both LPNO and LYNO was measured under applied magnetic fields up to 14 T with ***H***//***c*** and ***H***//***ab***, as also shown in Fig. 2 and Extended Data Fig. 2. For both systems, the applied magnetic field gradually suppresses the resistivity downturn upon increasing external fields as expected for a superconductor. For LPNO in Figs. 2(a) and 2(b), ***H***//***c*** suppresses the resistivity downturn much more effectively than ***H***//***ab***, reflecting strong anisotropy, which we further corroborate below. Similar anisotropy is seen in LYNO, but in contrast, the low-resistance state in LYNO persists up to 14 T for both magnetic-field orientations at low temperatures, consistent with its larger zero-field transition temperature (Extended Data Figs. 2(a) and 2(b)). The finite residual resistance in both systems, especially in LYNO, is likely due to the small superconducting volume fraction (see below).

Measurements of the isothermal magnetoresistance in magnetic field orientations with ***H***//***c*** and ***H***//***ab*** are shown for LPNO ($x$=0.55) and LYNO ($x$=0.11) in Figs. 2(c) and 2(d) and Extended Data Figs. 2(c) and 2(d), respectively. The magnetoresistance curves exhibit an abrupt increase with increasing magnetic field, consistent with a superconducting to normal transition, and this behavior gradually weakens as the temperature increases, eventually disappearing at $T_C$. For LPNO, ***H***//***c*** efficiently suppresses the superconductivity, giving rise to sharp, cusp-like features at intermediate temperatures. In contrast, ***H***//***ab*** is much less effective at suppressing the superconductivity, resulting in broader, dip-shaped and plateau-like features. For LYNO, the magnetoresistance curves show a less sharp behavior. Indeed, at low temperatures, the effect of the magnetic field is small, presumably because the critical field required to fully suppress the superconductivity exceeds the maximum applied magnetic field. We surmise that the anisotropy is not as clear for LYNO because the field scale is larger, but the evidence points to the conclusion that LYNO is less anisotropic than LPNO.

The magneto-transport data shown in Fig. 2 and Extended Data Fig. 2 allows for determination of the superconducting *H-T* phase diagrams, $\mu_0 H_{c2}(T)$, of LPNO and LYNO. Here, we define the transition temperature (field) as the 50% point of the main resistive drops. The resulting phase diagrams are shown in Extended Data Figs. 3(a) and 3(b). For LPNO, we observe a distinctly different behavior for the in-plane and out-of-plane orientations. For ***H***//***c***, the $\mu_0 H_{c2}$ curve displays some upwards curvature at temperatures slightly below $T_C$, which gives way to the conventional linear dependence at lower temperatures. The upwards curvature could be a signature of an inhomogeneous distribution of $T_C$-values or possibly a multi-gap superconducting state[42,43] In contrast, the $\mu_0 H_{c2}$ curve with ***H***//***ab*** displays pronounced downwards curvature which near $T_C$ is well described by a square-root variation as indicated by the green fit line. This phenomenology is characteristic of a thin film superconductor for which the phase boundaries are given within the Ginzburg-Landau formalism as[44] $\mu_0 H_{c2}^{c} = \phi_0/2\pi\xi_{GL}^2\,(1 - T/T_c)$ and $\mu_0 H_{c2}^{ab} = \sqrt{12}\phi_0/2\pi\xi_{GL} d\,\sqrt{1 - T/T_c}$. Here, $\phi_0$ is the flux quantum, $\xi_{GL}$ the Ginzburg-Landau coherence length and $d$ the thickness of the superconducting layer. With an extrapolated value $\mu_0 H_{c2}^{c}(0) \simeq 9$ T , corresponding to $\xi_{GL} \simeq 6\,nm$, we deduce from the square-root fit a layer thickness of $d \approx 10\,nm$. At low temperatures, the measured in-plane upper critical fields clearly fall short of the square-root dependence. This behavior may arise from paramagnetic limiting as the conventional Pauli paramagnetic limit,[45,46] $\mu_0 H_P = 1.84\ T_C$, amounts to ~14 T. $\mu_0 H_{c2}$ of LYNO (Extended Data Fig. 3(b)) does not display a distinct 2D-behavior but is rather more typical of an anisotropic superconductor with an anisotropy of ~3. We suggest, though, that the analysis of the critical field becomes increasingly error prone in the limit of exceedingly low volume fractions (see below).

Temperature-dependent AC susceptibility, presented as $4\pi\chi$ (in CGS units) in Fig. 3, was measured in both LPNO ($x$=0.55) and LYNO ($x$=0.11) with $\boldsymbol{H}//\boldsymbol{c}$ and $\boldsymbol{H}//\boldsymbol{ab}$. When $\boldsymbol{H}//\boldsymbol{c}$, breaks in the slopes of $4\pi\chi$ occur for both materials at the $T_C$ identified from transport, signaling the onset of a weak diamagnetic contribution to the susceptibility. As the DC bias field is increased, the diamagnetic response is progressively smeared out as expected for a superconductor. $4\pi\chi$ provides an estimate of the superconducting volume fraction with zero DC bias, which is ≈2% and ~0.03% for LPNO and LYNO, respectively. Although the estimated diamagnetic volume fractions are small, quantitative determination is subject to considerable uncertainty. In addition to all the well-known caveats for such estimates[47], an additional complication in this case is the presence of ferromagnetic nickel which arises from the reduction of nickel oxide impurities[48] that formed during the floating zone growth and hyodrogen reduction. It is important to note that we did not detect a diamagnetic response in the susceptibility with $\boldsymbol{H}//\boldsymbol{ab}$. This observation is consistent with the 2D nature of the superconductivity.

Current–voltage ($I$-$V$) characteristics of LPNO ($x$=0.55) measured at 2 K under applied magnetic fields up to 14 T with $\boldsymbol{H}//\boldsymbol{c}$ are shown in Fig. 4(a) and at zero-field over the temperature range of 2–10 K in Fig. 4(c). The corresponding differential resistance ($\mathrm{d}V/\mathrm{d}I$) characteristics are shown in Figs. 4(b) and 4(d), respectively. A nonlinear response is observed in the superconducting state, which evidences the critical current threshold. An Ohmic background can be attributed to the finite resistivity of the non-superconducting regions. As the applied magnetic field or temperature increases, the nonlinearity in the *I*-*V* characteristics is suppressed, as expected for superconductivity. The corresponding *I*-*V* and d*V*/d*I* characteristics of LYNO are displayed in Figs. 4(e), (g), (f), and (h), respectively. In Fig. 4(e), all *I*-*V* curves overlap in the low-current regime with linear response and begin to diverge at $I \gtrsim 1$ mA with increasing magnetic field, indicating that the nonlinear contribution from the superconducting state persists up to 14 T while being progressively suppressed by the applied magnetic field. As shown in Fig. 4(g), the observed nonlinearity is gradually smeared out with increasing temperature similar to that observed in LPNO. Critical current behavior is readily apparent in d*V*/d*I*, reflecting the underlying nonlinearity in the *I*-*V* characteristics. In both LPNO and LYNO, the two characteristic peaks in d*V*/d*I*, which serve as a measure of the critical current, disappear at approximately 7 K and 15 K, respectively, coinciding with the $T_C$ measured in resistivity.

**Discussion**

Here we summarize the findings that we consider key towards understanding the non-percolative superconductivity in LPNO and LYNO. First, superconductivity emerges only in samples found at the phase boundary between the metallic and charge/spin-stripe insulating phases in both the LPNO and LYNO series, a region of phase co-existence. The fact that superconductivity appears at this phase boundary in two distinct systems indicates its importance to the phenomenology. In LPNO, superconductivity was observed over a range of *x* of approximately 0.52 to 0.6, whereas in LYNO it was observed near 0.11. First, in all crystals of LPNO that exhibit superconductivity, the nearly same $T_C$ is observed; the same holds for LYNO (albeit with a higher $T_C$). A possible explanation is that compositional inhomogeneities within crystals lead to a situation where optimally tuned $T_C$ (as a function of $x$) is always observed for each series, similar to the case of $La_3Ni_2O_{7-\delta}$ (as a function of $\delta$)[49]. Second, anisotropic magnetoresistance and AC susceptibility measurements reveal that the superconducting fraction is anisotropic with respect to the applied magnetic-field direction. Specifically, the rapid suppression of the superconducting state for ***H***//***c***, together with the absence of a detectable diamagnetic response for ***H***//***ab*** in AC susceptibility measurements, is consistent with a 2D superconductor. Third, the observed diamagnetic shielding fraction remains only a few percent for LPNO and less than 1% for LYNO, consistent with the presence of non-percolative superconducting regions. In the following paragraphs, we discuss several potential origins of superconductivity that we consider to be most plausible and compatible with our observations and rule out two alternative explanations.

**Surface superconductivity.** We considered whether the superconductivity could be a surface effect. To check this, we investigated samples with different surface areas (i.e. crushed crystals vs whole crystals) but found that the volume fraction did not improve with larger surface areas. We therefore excluded surface superconductivity as the origin.

**Oxygen inhomogeneities.** Hydrogen reduction from the 4-3-10 phase to the 4-3-8 phase is a complex process that occurs through intermediate phases such as 4-3-9[50,51]. Excess oxygen anions can form ordered patterns that may be inhomogeneously distributed throughout the crystal as the reduction proceeds. Thus, we considered whether low volume fraction(s) of a 4-3-(8+δ), 4-3-(9+δ), and/or 4-3-(10+δ) phases were present that might be superconducting. To test this possibility, we performed AC susceptibility measurements on a single crystal as it was stepwise reduced. The essential finding from our experiments was that the shielding fraction increased as the reduction time was increased; however, once a threshold was reached, further annealing in reducing conditions led to no further increase in shielding fraction. More details about these experiments are in SI Section II. Our view is that continued reduction favors the 4-3-8 phase at the expense of the intermediate phases, and thus disfavors assignment of superconductivity to intermediate reduced phases. Moreover, given that both $Pr_4Ni_3O_{10}$ and $La_4Ni_3O_{10}$ reduce to $Pr_4Ni_3O_8$ and $La_4Ni_3O_8$, respectively, under the same conditions, we consider it unlikely that the reduction process would occur differently in their solid solution. There is no obvious reason for special oxygen-ordered structures, which form at high temperatures, to possess a compositional dependence that is correlated to the bulk low temperature 4-3-8 phase diagram. We therefore exclude this possibility.

**Superconducting square-planar intergrowths**. Scanning transmission electron microscopy (STEM) imaging of a LPNO crystal ($x$=0.55) reveals a high density of defects, as shown in Figs. 5(a) and Extended Data Fig. 4. These defects can be classified into phase-separated regions and intergrowths (IG). The phase-separated regions do not exhibit any signature of the perovskite-type structure. They show polycrystallinity and amorphization with clear Ni segregation. On the other hand, we observe that IGs exhibit a perovskite-derived structure with various values of $n$ ($n$ = 7, 31, and 44). Figure 5(b) shows a STEM-low-angle annular dark-field (LAADF) image for an IG with $n$ = 31. The bright contrast in the IG points towards a high epitaxial strain. Moreover, these IGs are incoherently strained and have a high density of planar defects and dislocations (highlighted as blue boxes in Figs. 5(b) and 5(c)). Figures 5(c) and 5(d) show the atomic-resolution STEM- high-angle annular dark-field (HAADF) images of IGs with $n$ = 44 and 7, respectively. To qualitatively estimate the nature of the IGs (perovskite vs reduced), we compared the average R-R distance along the $c$-axis ($d_{R\text{-}R}$) to that for the underlying trilayer structure, which is 3.56 Å. For the IGs with $n$ = 44, 31, and 7, the total IG layer thicknesses are 15.897 nm, 11.202 nm and 2.77 nm, respectively. These correspond to average $d_{R\text{-}R}$ of 3.61 Å, 3.61 Å, and 3.96 Å, respectively, for $n$ = 44, 31, and 7. These values indicate that the $n$=44 and 31 IGs were reduced (i.e., square-planar), whereas the $n$=7 IG was an under-reduced 7-layer perovskite phase. Thus, during the initial floating-zone growth, perovskite IGs with $n$>3 (some with $n$>>3) are formed. Hydrogen reduction reduces most of these perovskite IGs to square-planar. The square-planar IGs with $n$>>3 approximate slabs of the infinite-layer 1-1-2 phase. Strain in these IG regions may arise due to coherent clamping of the IGs to the main $n$=3 matrix, similar to related observations in 1-1-2[52] and 4-3-8 films.[53] Thus, they represent a potential source of superconductivity since strained 1-1-2 films are well-known to exhibit superconductivity, as discussed above. In fact, the superconducting layer thickness estimated from the in-plane $\mu_0 H_{c2}$ curve of $d$≈10 nm falls in the range of observed intergrowth thickness. Additionally, the strain on these IGs may be significantly temperature-dependent, particularly given the evidence that crystals that possess superconducting regions also possess regions that undergo the 1st-order insulator-metal transition at $T_{DW}$>$T_C$. We expect that the insulating regions undergo changes in local atomic distances at $T_{DW}$ since the insulator-metal transition is accompanied by changes in the lattice constants at $T_{DW}$ in $La_4Ni_3O_8$[35]. We expect that strain effects are amplified for LYNO given the smaller atomic radius of $Y^{3+}$ as compared to $Pr^{3+}$.

An open question with the IG explanation is the compositional dependence of the superconductivity. Specifically, it is unclear why IGs of higher order phases should have compositional-based properties that are correlated to the underlying bulk metal-insulator phase boundary of the 4-3-8 matrix in which they reside.

**Interface superconductivity.** Given the co-existence of both metallic and insulating regions of the 4-3-8 structure along with various IGs, it is possible that the interfaces between different mesoscopic structures possess the right combination of strain and charge carrier concentration to yield superconductivity. As an example, an interface between 1-1-2 and 4-3-8 resembles that of a superconducting interface formed between an un-doped cuprate and a non-superconducting over-doped cuprate [54].We suggest that deliberate synthesis of epitaxial interfaces (e.g. 1-1-2 on 4-3-8) may help establish whether this could be a viable explanation.

**Competing Superconductivity and Stripes.** As discussed above, superconductivity frequently arises when a charge or spin-ordered state is suppressed towards $T \rightarrow 0$ via a tuning variable, as described schematically in Fig. 1(a). Indeed, there is an emerging picture of the progression of phase behavior in the related bilayer Ruddlesden-Popper $La_3Ni_2O_7$ in which oxygen stoichiometry or strain tunes an intertwined CDW/SDW first to a superconducting state and then to a non-superconducting metallic state[49,55–57]. In LPNO and LYNO, the phase competition between charge-stripe insulating and metallic regions is expected to lead to progressively smaller and more isolated 'droplets' of the stripe phase embedded in the percolated metallic matrix as seen, for instance, in manganites[58]. At the phase boundary itself, these stripe regions will be fragile. The most straightforward pathway upon decreasing temperature would be for the droplets to simply transform into the metallic phase. A more intriguing possibility, however, is that in this $T$-$x$ regime correlated-metal, stripe, and superconducting states are close enough in free energy that the superconductor phase can nucleate locally at the boundary, leading to filamentary superconductivity. In this case, the superconducting volume fraction will fundamentally be limited by the complex morphology of the phase interface. Similar phenomenology has been argued in $Ba_{1-x}Sr_xNi_2As_2$[59] and 1T-$TiSe_2$[60], while theoretical models have been constructed in which superconductivity competes with and nucleates at CDW boundaries[61]. Earlier work by Kivelson, Fradkin and Emery[62] proposed superconductivity to exist in a smectic phase between an insulating stripe state and a non-superconducting metallic phase. In that context, our transport and AC susceptibility data are reminiscent of those of $La_{1.875}Ba_{0.125}CuO_4$ where stripe and superconducting order coexist[63]. Tests of this proposal could include pressure tuning to suppress the stripe phase or direct interrogation of the superconductor via scanning probe spectroscopy.

Finally, we address the commonality of the LYNO and LPNO phase diagrams, which are shown in Figs. 6(a) and 6(b). LYNO evolves much more quickly with $x$ than LPNO because of the smaller $Y^{3+}$ ions compared to $Pr^{3+}$. Attempts to correlate the insulator-metal transition temperatures with lattice constants, Ni-O bond lengths or bond angles failed. We also attempted to parameterize the transition temperatures by the total unit-cell volume (Fig. 6(c)). Although $T_{DW}$ for both exhibit similar trends with unit-cell volume, they do not coincide. Refinements of single-crystal X-ray diffraction data (see SI section I for results of XRD refinements) indicated that $Y^{3+}$ exclusively substituted into the $La^{3+}$ sites within the fluorite-like spacer layers. We therefore attempted to parameterize the transition temperatures by the fluorite-like spacer layer volume. Remarkable overlap of the two re-parameterized phase diagrams (Fig. 6(d)) indicated that the fluorite-like spacer layer volume unifies the two series as a common tuning parameter. From size arguments, we anticipate that Pr substitutes randomly in the two rare earth sites, explaining the more rapid $x$ dependence of $T_C$ for Y than Pr. However, the similar Z of La and Pr precludes verifying this assertion by x-ray diffraction. We note that the changes in the fluorite-like spacer layer volume are primarily dictated by its thickness. Thus, the thickness/volume of these fluorite-like spacer layers, which is controlled sterically, allows for tuning of the nickelate electronic phase diagram without aliovalent substitution.

## Conclusion

In summary, we have presented evidence for non-percolative, filamentary superconductivity in single crystals of LYNO and LPNO. Magneto-transport and susceptibility measurements reveal that the superconductivity is two-dimensional in nature. Importantly, superconductivity is confined to samples in the compositional phase space where charge carrier delocalization and charge/spin stripes are delicately balanced, leading to electronic inhomogeneity, mesoscopic phase segregation, and eventually a percolative insulator-metal transition. A plausible origin of the observed filamentary superconductivity is nucleation of the superconducting phase at the boundary between these mesoscopic phases. Notably, the electronic phase competition within Ni-O multilayers can be tuned by adjusting the thickness of the spacers in between the multilayers, offering a materials design strategy for creating and investigating related phases for ambient pressure superconductivity. We point out that bulk single crystals of reduced trilayer nickelates inevitably harbor intergrowth defects of infinite layer-like nickelate slabs, which offer an alternative explanation for the source of the low volume fraction superconducting phase. At present, the data at hand does not allow us to unambiguously distinguish between these possibilities. What is clear, however, is that mixed phase systems at metal-insulator boundaries, particularly those involving charge/spin-stripes, offer a provocative opportunity space for further exploration of ambient pressure superconductivity in nickelates.

Future studies combining spatially resolved probes will be essential for unraveling the intrinsic physics of the superconductivity in these materials from the materials science perspective. In particular, atomic-resolution measurements of oxygen stoichiometry across the phase boundary, local strain mapping, and nanoscale spectroscopic techniques that are capable of directly imaging superconducting regions, would provide critical insight into the microscopic origin of the superconductivity. Data from such studies would help to establish whether the observed superconductivity originates from oxygen stoichiometry, strain-induced effects, or an intrinsic pairing mechanism associated with the metal-insulator phase boundary, as seen in other unconventional superconductors[64].

**Author Contributions**

Crystal Growth: X.C. and H.Z.; Crystal Reduction: X.C., H.-B.A., Y.Z., and H.Z.; X-Ray Analysis: X.C., Y.L., and J.F.M.; Experimental Design: X.C., H.-B.A., J.F.M., U.W., M.R.N., and D.P.; Magneto-transport Measurements: H.-B.A., X.C., U.W., and R.C.; Susceptibility: H.-B.A. and U.W.; Microscopy: A.S.T. and R.F.K.; Manuscript Writing: H.-B.A., D.P., M.R.N., U.W., and J.F.M.

**Acknowledgement**

Work in the Materials Science Division of Argonne National Laboratory (crystal growth, X-ray diffraction, (magneto)transport measurements, susceptibility measurements, analysis, manuscript writing) was supported by the U.S. Department of Energy, Office of Science, Basic Energy Sciences, Materials Sciences and Engineering Division. A.S.T and R.F.K. were supported by the Office of Basic Energy Sciences, U.S. Department of Energy, award number DE-SC0025396. This work made use of the ThermoFisher Helios 5CX (cryo) FIB-SEM instrument in the Electron Microscopy Core of UIC's Research Resources Center, which received support from UIC, Northwestern University and ARO (W911NF2110052). Acquisition of the UIC JEOL ARM200CF was supported by an MRI-R2 grant from the National Science Foundation (DMR-0959470). The Gatan Continuum GIF acquisition at UIC was supported by an MRI grant from the National Science Foundation (DMR-1626065). Work performed at the Center for Nanoscale

Materials, a U.S. Department of Energy Office of Science User Facility, was supported by the U.S. DOE, Office of Basic Energy Sciences, under Contract No. DE-AC02-06CH11357.

## Methods

### Crystal growth and characterization

$(La_{1-x}Pr_x)_4Ni_3O_{10}$ and $(La_{1-x}Y_x)_4Ni_3O_{10}$ (4-3-10) crystals were grown under high oxygen pressures (25 to 140 bar) in a ScIDre GmbH HKZ floating zone furnace. They were then transformed into LPNO and LYNO by reduction in a 3.5% $H_2$ gas balanced with Ar at $T$≈340 °C for ≈4 days. Because the solid solution in LPNO exists for all values of substitution between $La_4Ni_3O_8$ and $Pr_4Ni_3O_8$, we take the value of $x$ from the ratio of starting reagents. On the other hand, for LYNO our crystal growth reached a solubility limit of $x$~0.15, necessitating a need to determine compositions from single-crystal X-ray refinements, which were possible due to the significant difference in scattering cross-sections of $Y^{3+}$ and $La^{3+}$ ions. The X-ray refinements (SI Section I) were made from single crystal diffraction data taken with a Bruker APEX2 area detector and Mo Kα radiation (λ = 0.71073 Å). AC susceptibility was measured with a Quantum Design MPMS3 SQUID magnetometer under a 1 Oe magnitude field oscillating at 1 kHz. Magneto-transport measurements were performed in a 14-T Quantum Design PPMS.

### Electron Microscopy

The cross-sectional samples for STEM experiments were prepared using a Thermo Fischer Scientific Helios 5 CX focused-ion beam (FIB)/scanning electron microscope (SEM) DualBeam system at the University of Illinois Chicago. The final step of the lamella polishing was performed with an ion beam energy of 1 kV in order to minimize surface damage. STEM experiments were conducted at the University of Illinois Chicago using a JEOL JEM-ARM200CF microscope, which was operated at 200 kV. The microscope is equipped with a CEOS aberration corrector and cold-field emission source. HADDF and LAADF imaging were performed using an electron probe convergence semi-angle of 30 mrad. The inner and outer collection angles for HAADF and LAADF imaging were set to 90 – 370 mrad and 40 – 160 mrad, respectively. The high-quality atomic-resolution images were sequentially acquired (~10 – 15 frames) and were subsequently aligned and integrated.

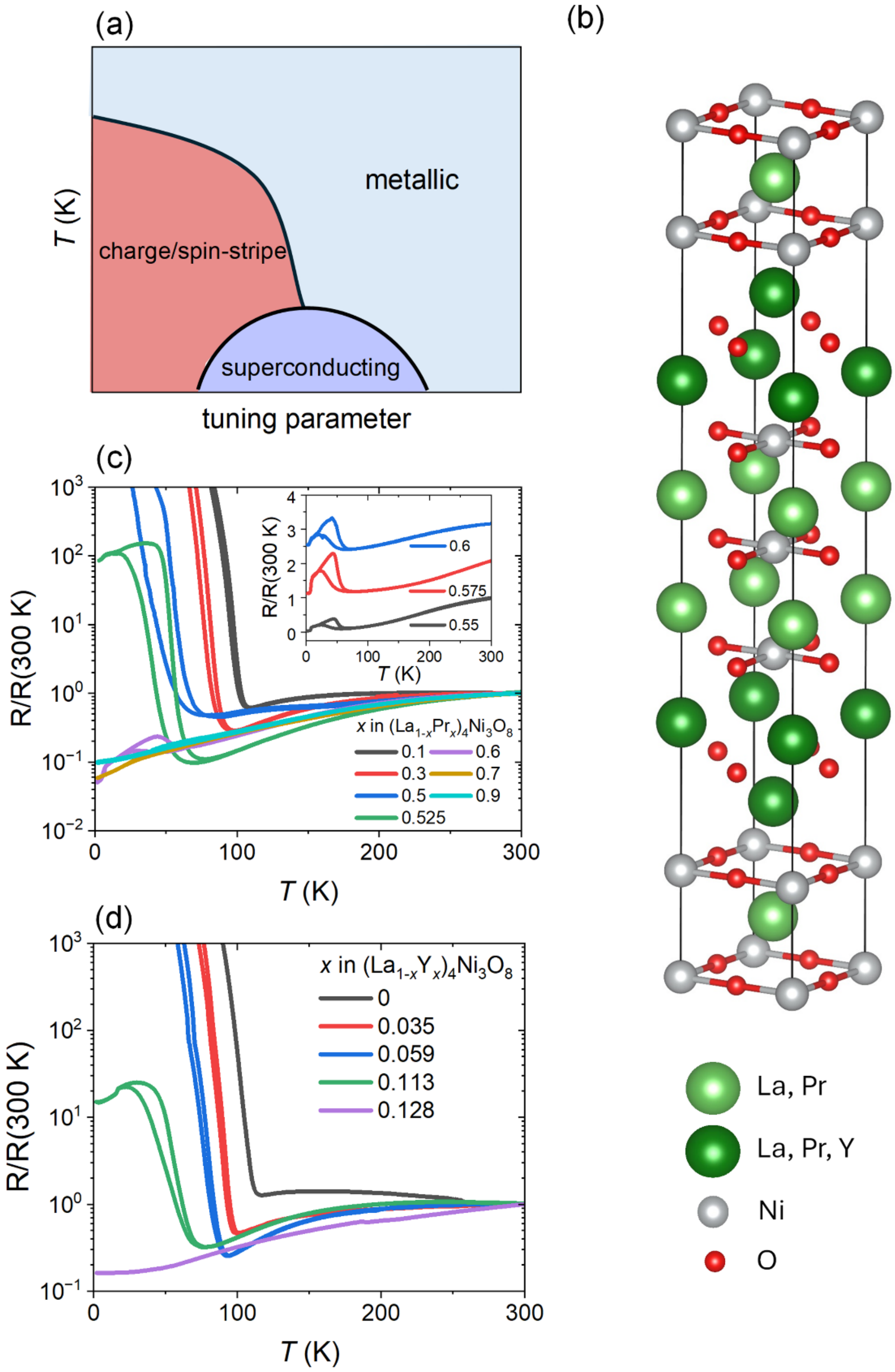

Fig. 1: (a) Generic phase diagram of a tuning parameter suppressing a density wave leading to superconductivity. (b) Trilayer $R_4Ni_3O_8$ structure. The R sites are distinguished by two colors of green to separate those from the fluorite-like spacer layers from those inside the trilayers. (c, d) Normalized resistance versus temperature for LPNO and LYNO for different values of $x$. The inset in (c) shows the data over a narrower range of $x$, from 0.55 to 0.6.

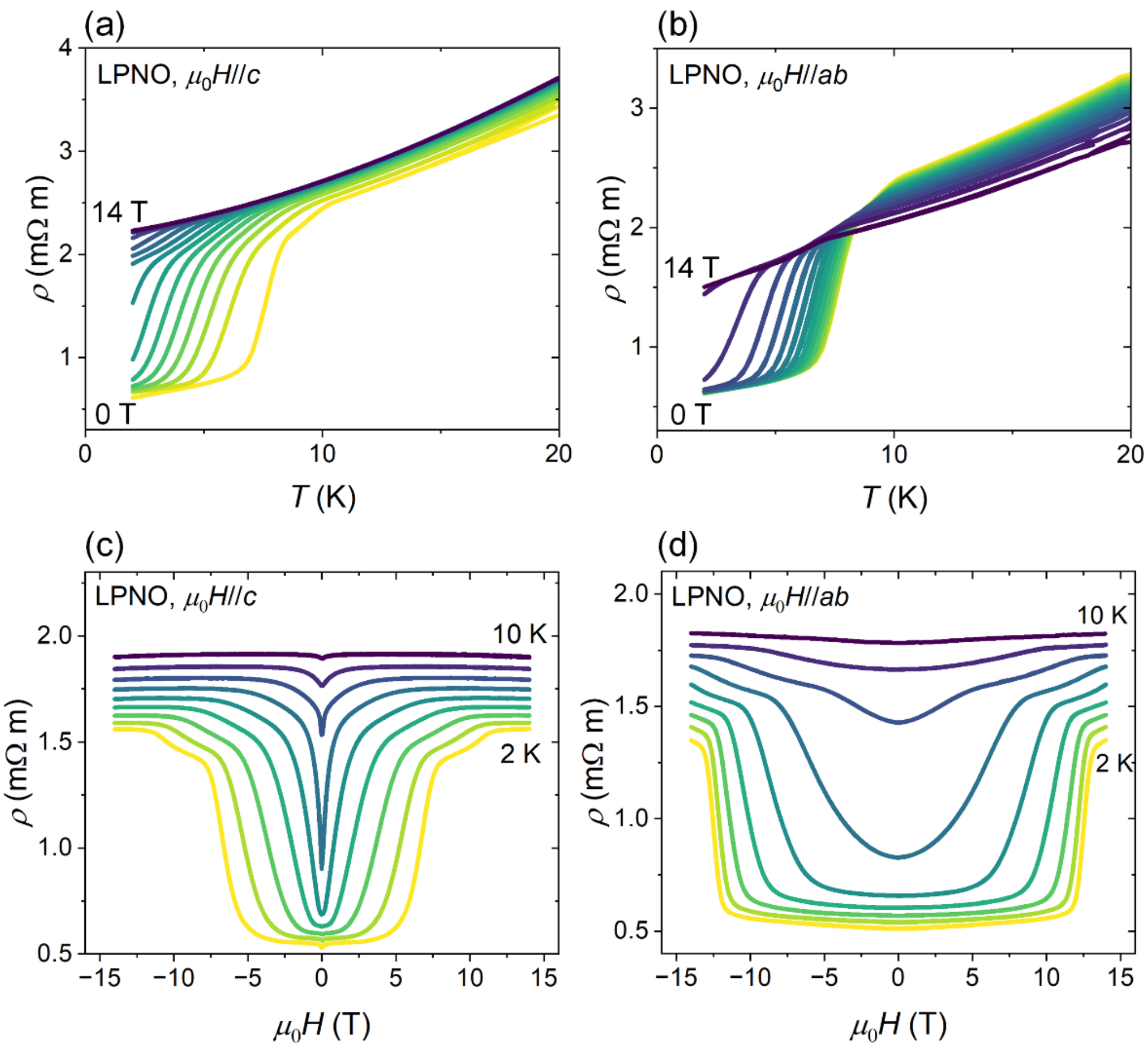

Fig. 2: Resistivity as a function of temperature for LPNO ($x$=0.55) measured with various magnetic fields applied parallel to the *c*-axis (a) and parallel to the *ab*-plane (b), and isothermal field sweeps of the resistivity measured with magnetic field applied parallel to the *c*-axis (c) and parallel to the *ab*-plane (d).

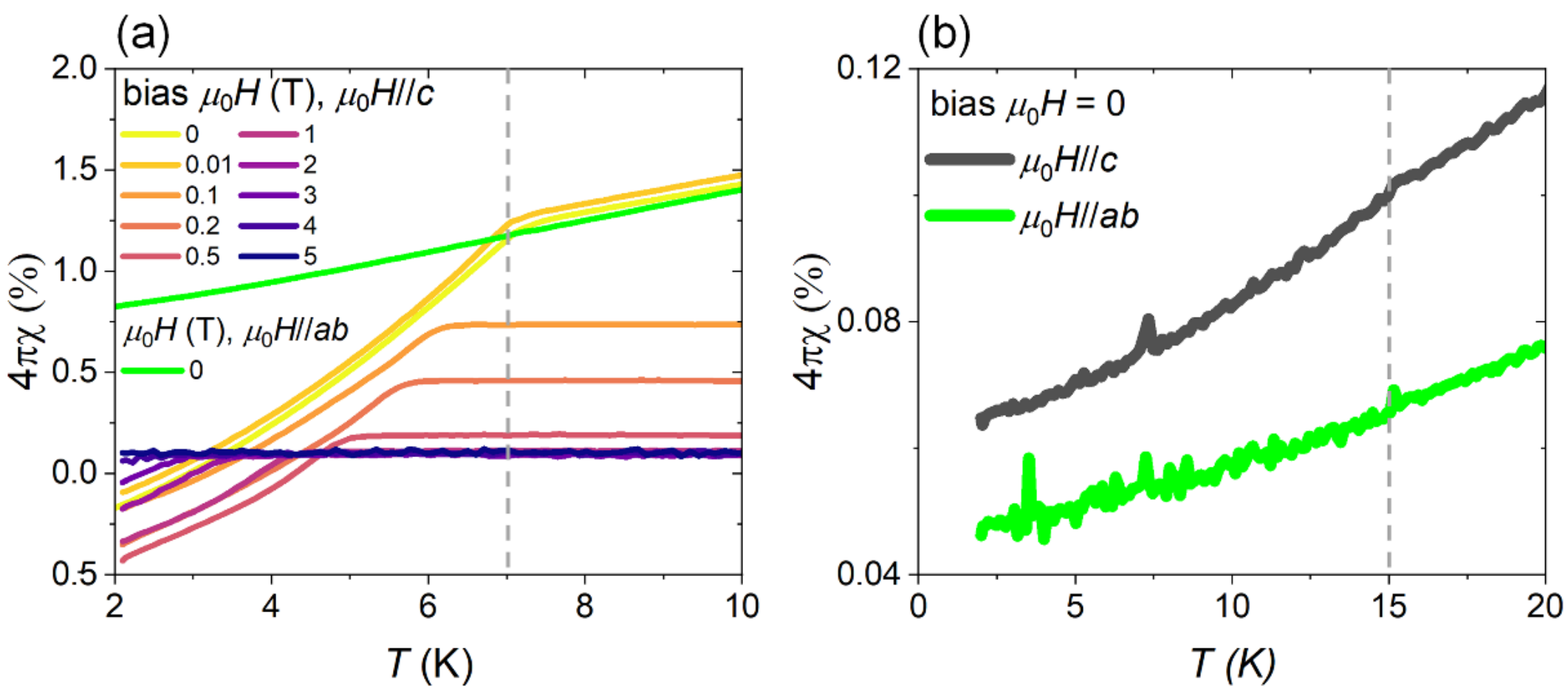

Fig. 3: χ' from AC susceptibility measurements of (a) LPNO ($x$=0.55) and (b) LYNO ($x$=0.11) with applied DC bias fields parallel to $c$-axis, as specified. Note the absence of a diamagnetic anomaly for the magnetic field applied parallel to $ab$-plane (green curves). The vertical dashed line indicates the zero-field resistive $T_C$.

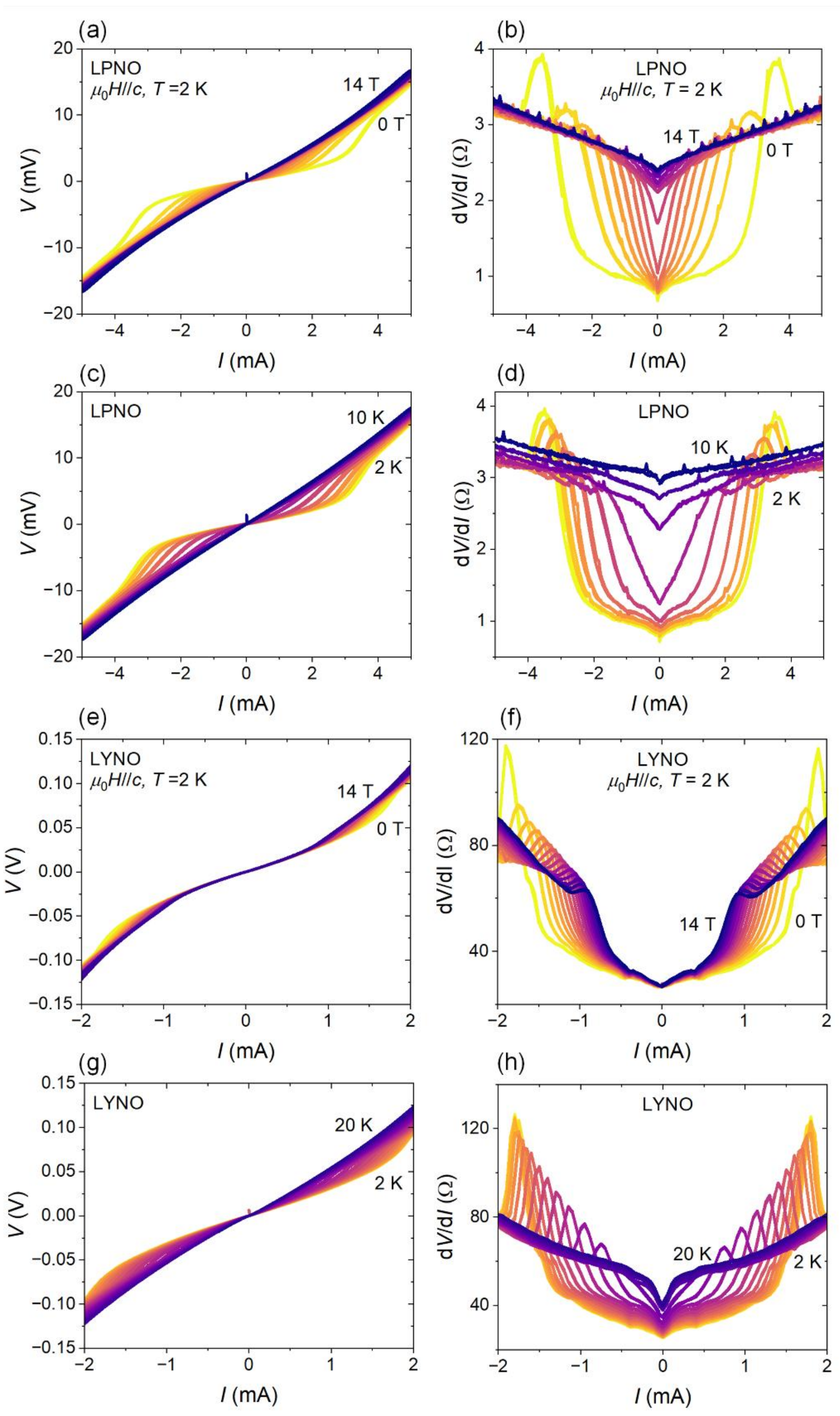

Fig. 4: (a) *I*-*V* characteristics of LPNO ($x$=0.55) at $T$=2 K measured with various magnetic fields applied parallel to the $c$-axis. (b) Differential resistance (d$V$/d$I$) for data shown in (a). (c) *I*-*V* characteristics of LPNO ($x$=0.55) at zero-field for various temperatures. (d) Differential resistance for data shown in (c). (e-h) Corresponding plots to (a-d) for LYNO ($x$=0.11).

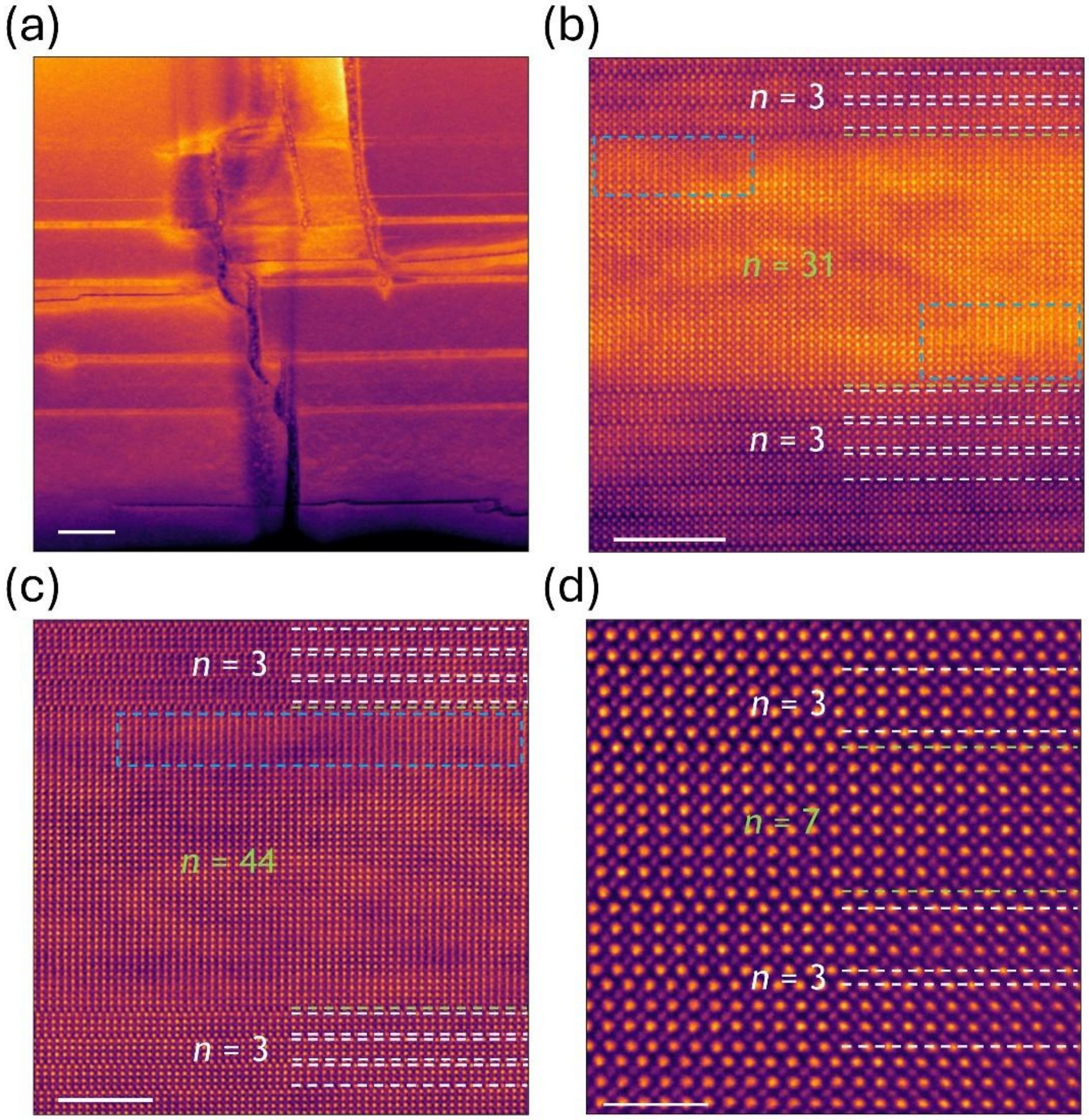

Fig. 5: STEM images for an LPNO crystal ($x$=0.55). (a) A wide field-of-view STEM-LAADF image showing a high density of defects. (b) STEM-LAADF image showing an incoherently strained intergrowth region with $n = 31$. STEM-HAADF images showing intergrowth regions with (c) $n = 44$ and (d) $n = 7$, respectively. The blue boxes in (b) and (c) highlight the defects associated with the incoherent strain in the intergrowth regions. Scale bars correspond to 100 nm for (a), 5 nm for (b) and (c), and 2 nm for (d).

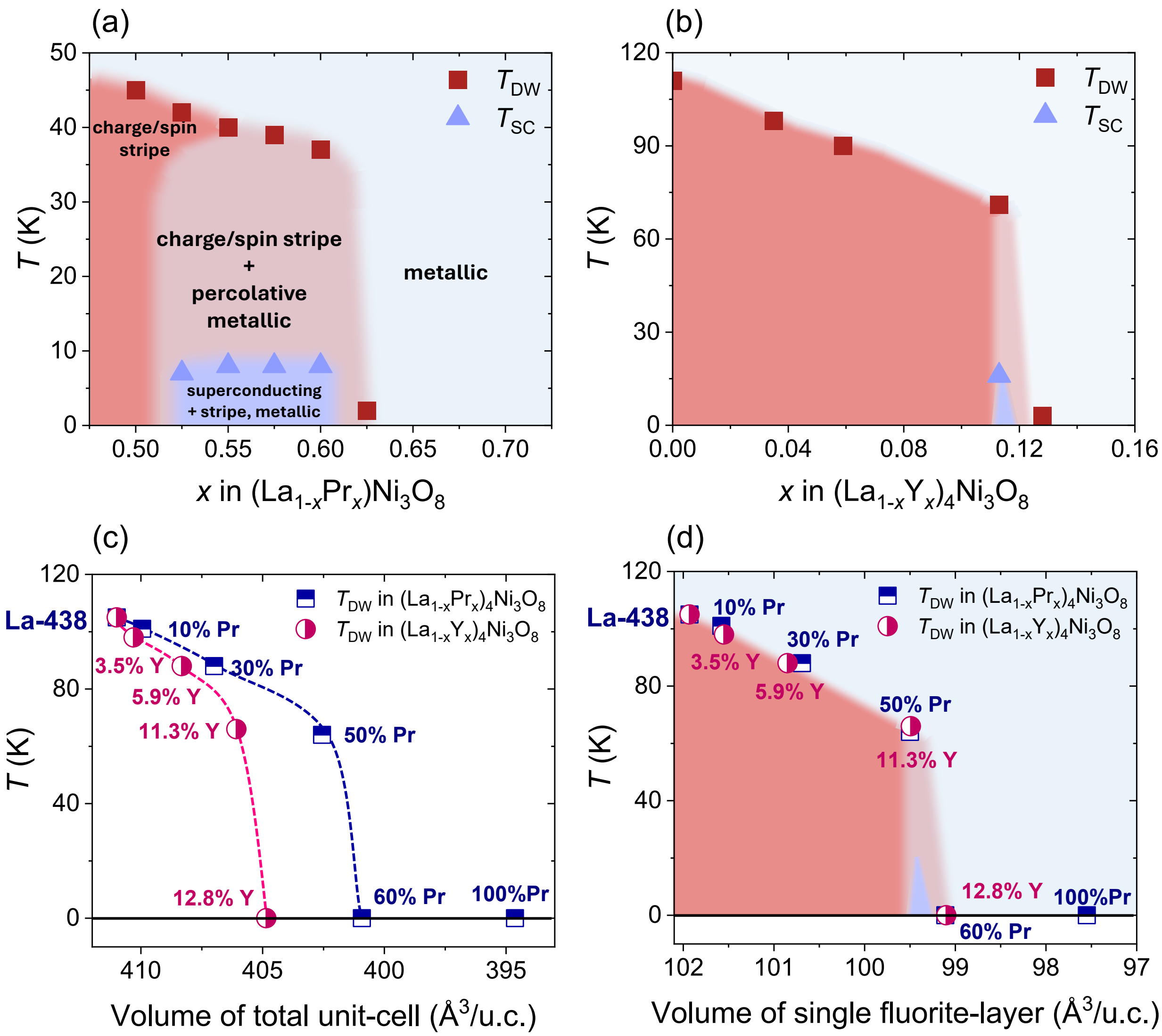

Fig. 6: Phase diagrams of LPNO (a) and LYNO (b) as a function of *x*. The coexistence of charge/spin stripes and metallic regions suggests the presence of a percolative metallic phase within the charge/spin stripe background. Common phase diagram for both LPNO and LYNO as a function of total unit cell (c) and re-parameterized as a function of the fluorite-like spacer layer volume (d). Each data point in phase diagram represents the onset temperature of the density-wave transition.

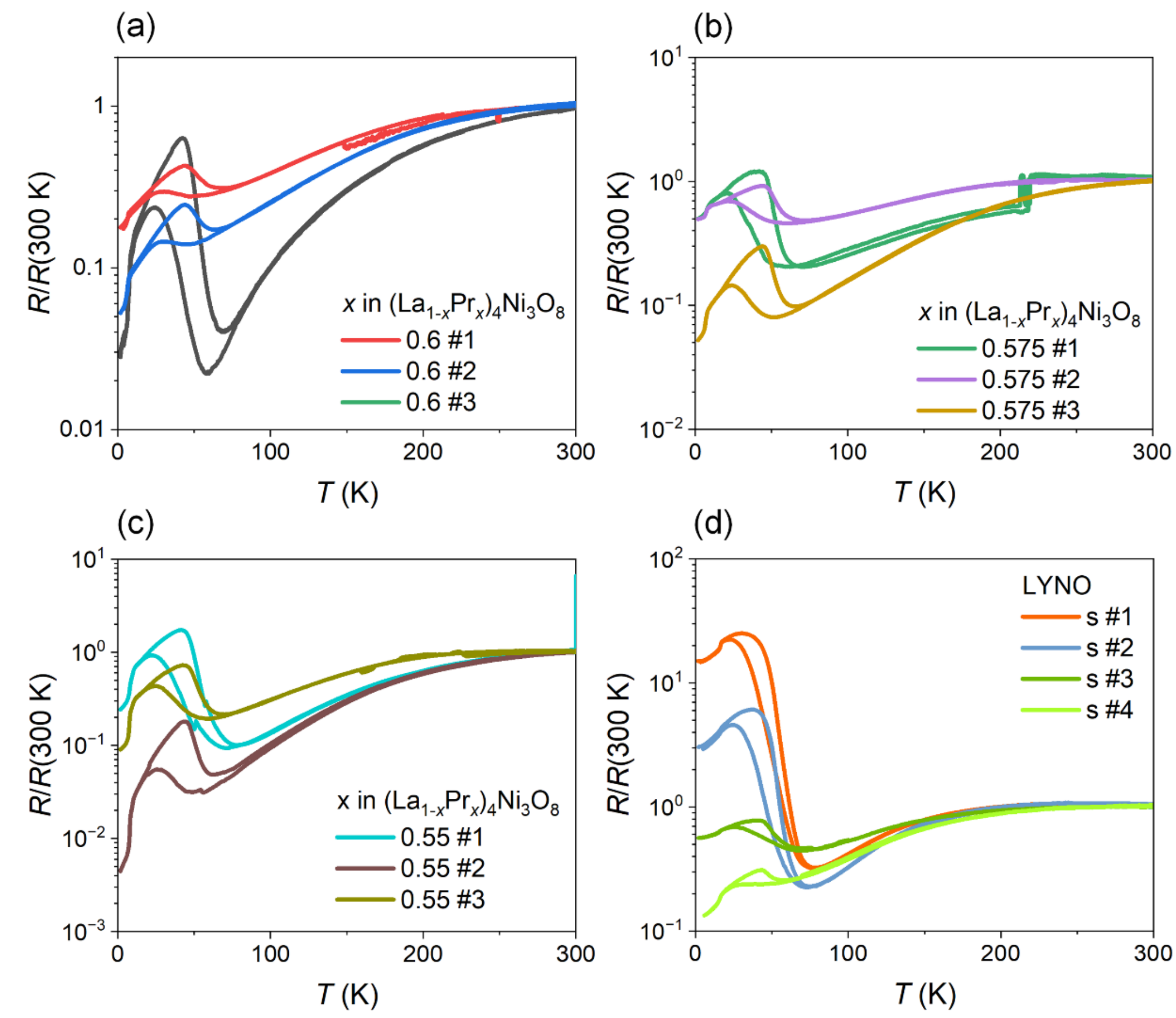

Extended Data Fig. 1: Normalized resistance versus temperature for several LPNO crystals over a narrower range of $x$, for 0.55 (a), 0.575 (b), and 0.6 (c), respectively. (d) Normalized resistance versus temperature for several LYNO crystals near $x$=0.11 that show a superconducting downturn.

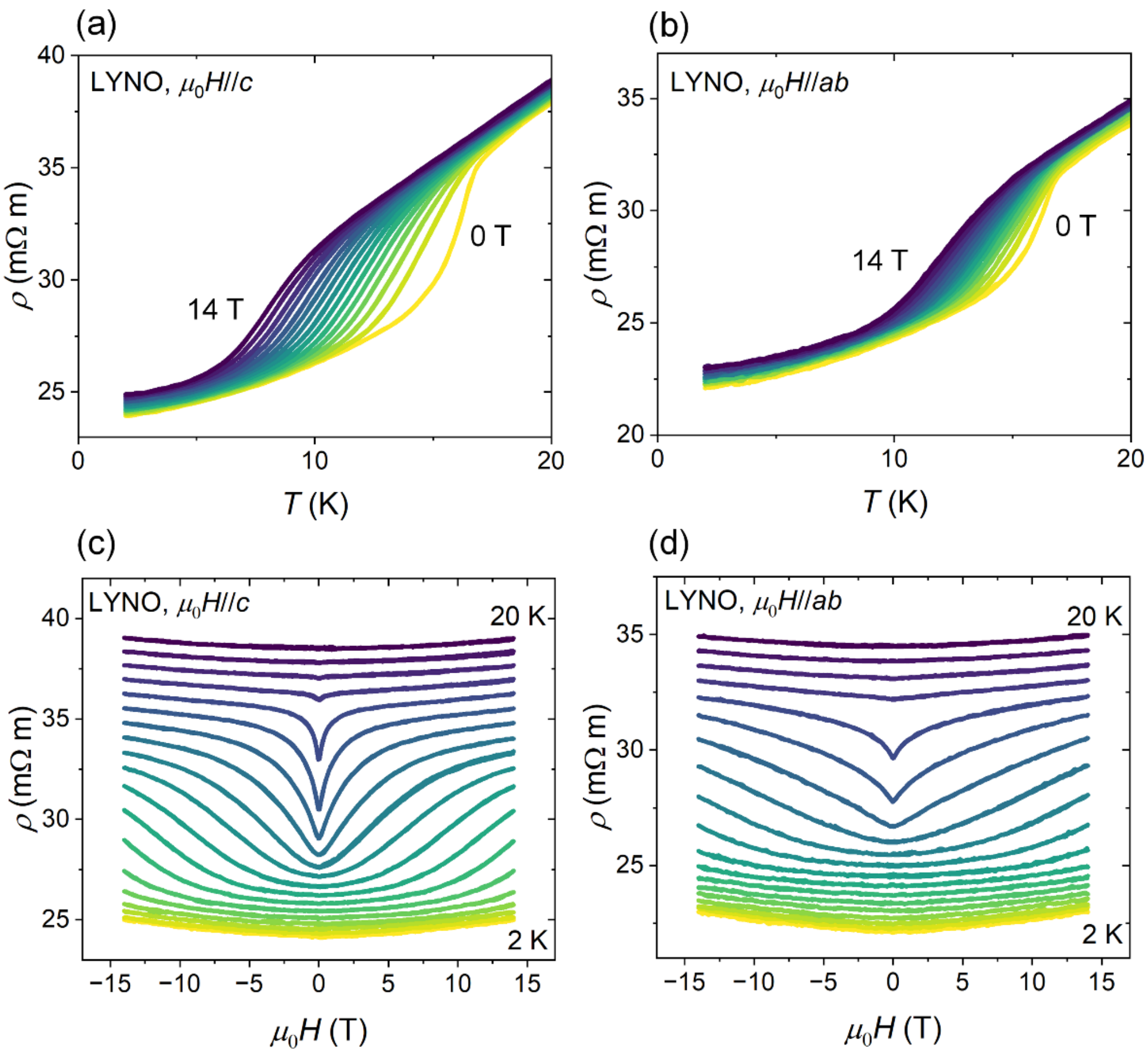

Extended Data Fig. 2: Resistivity as a function of temperature for LYNO ($x$=0.11) measured with various magnetic fields applied parallel to the *c*-axis (a) and parallel to the *ab*-plane (b), and isothermal field sweeps of the resistivity measured with magnetic field applied parallel to the *c*-axis (c) and parallel to the *ab*-plane (d).

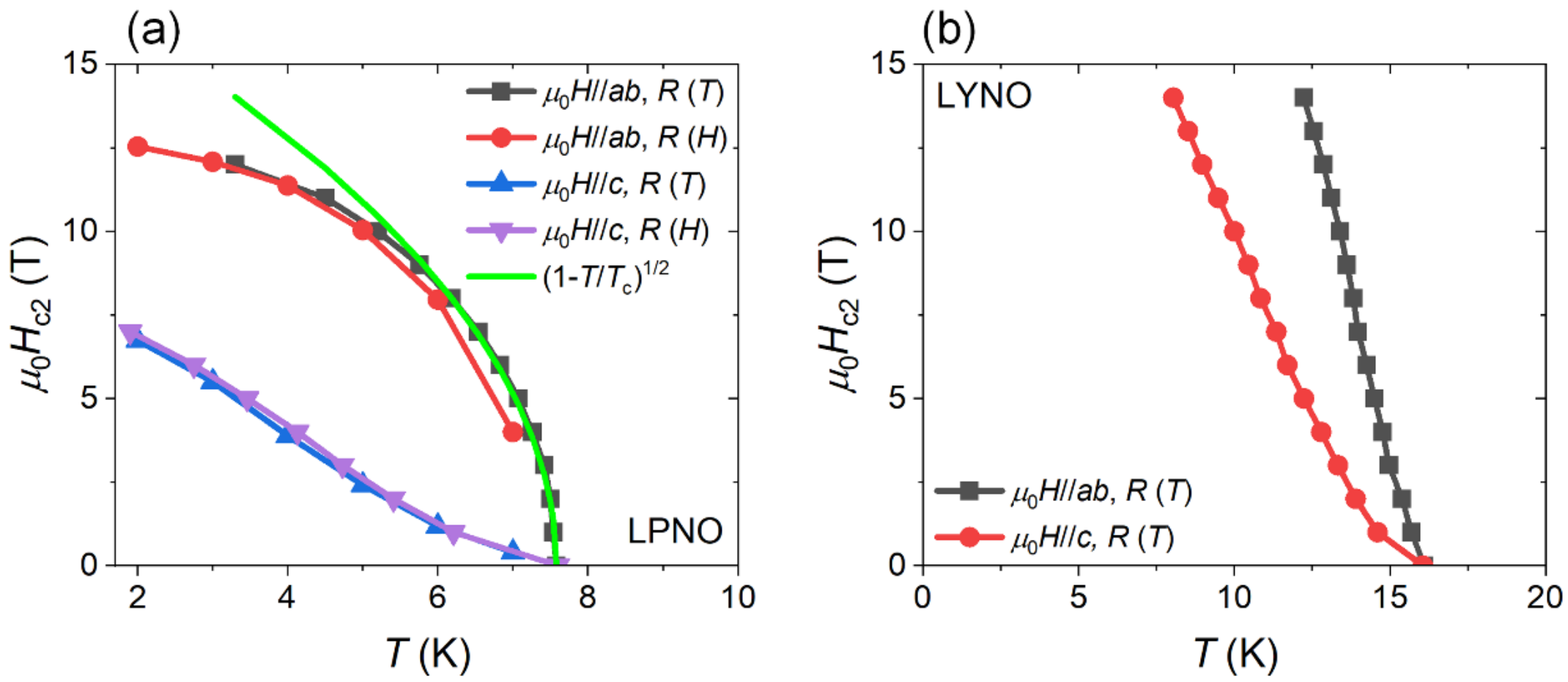

Extended Data Fig. 3: (a) Temperature dependence of the upper critical field, $\mu_0 H_{c2}$, of LPNO ($x$=0.55), measured with various magnetic fields applied parallel to $c$-axis and parallel to the $ab$-plane. The green curve is a fit for the square-root variation with $T$ of the in-plane $\mu_0 H_{c2}$. (b) $\mu_0 H_{c2}(T)$ of LYNO ($x$=0.11), measured with various magnetic fields applied parallel to the $c$-axis and parallel to the $ab$-plane.

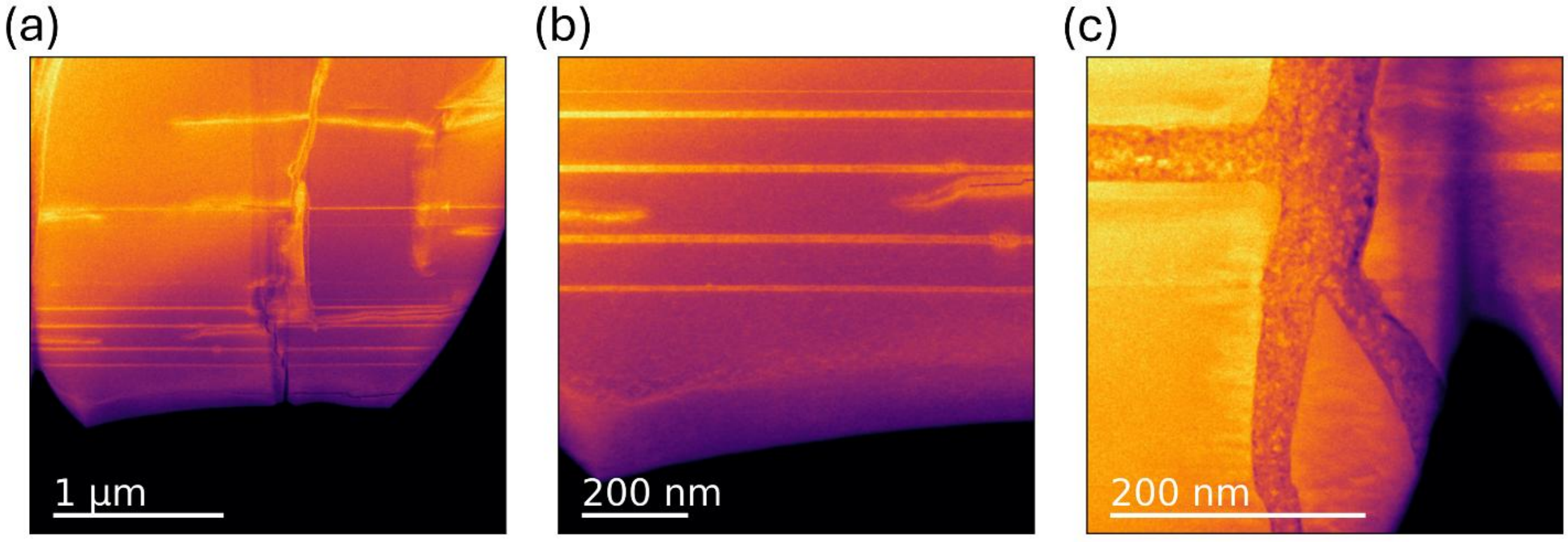

Extended Data Fig. 4: (a) STEM-LAADF image showing a high density of defects in LPNO ($x$=0.55). STEM-LAADF images showing (b) intergrowth and (c) phase-separated regions in an LPNO crystal (x=0.55).

Supplementary Information for

# Superconductivity at the metal-insulator phase boundary in a bulk nickelate at ambient pressure

Hyo-Bin Ahn[1,*], Xinglong Chen[1,*,#], Hong Zheng[1], Yu Zhang[1], Ramakanta Chapai[1,♠], Yu Li[1], Arashdeep S. Thind[2], Robert F. Klie[2], Michael R. Norman[1], Ulrich Welp[1], J.F. Mitchell[1], and Daniel Phelan

[1]Material Science Division, Argonne National Laboratory, Lemont, Illinois 60439, USA

[2]Department of Physics, University of Illinois Chicago, Chicago, IL 60607, USA

[*] These authors contributed equally to this work

[#] Current affiliation: School of Physics, Southeast University, Nanjing 211189, China

♠ Current affiliation: Department of Physics, Norfolk State University, Norfolk, VA 23504, USA

## I. Results of refinements of single crystal X-ray diffraction data for $(La_{1-x}Y_x)_4Ni_3O_8$.

| Crystal data and structure refinement for $La_{1-x}Y_xNi_3O_8$ | | | | | |
|---|---|---|---|---|---|
| Empirical formula | $La_4Ni_3O_8$ | $La_{3.859}Y_{0.141}Ni_3O_8$ | $La_{3.764}Y_{0.236}Ni_3O_8$ | $La_{3.514}Y_{0.452}Ni_3O_8$ | $La_{3.488}Y_{0.512}Ni_3O_8$ |
| Y content ($x$) | 0% | 3.5(6)% | 5.9(5)% | 11.3(5)% | 12.8(4)% |
| Space group | I4/mmm | I4/mmm | I4/mmm | I4/mmm | I4/mmm |
| a/Å | 3.9683(3) | 3.9683(4) | 3.9628(2) | 3.9589(3) | 3.9556(4) |
| c/Å | 26.099(3) | 26.055(3) | 26.0015(17) | 25.910(2) | 25.874(3) |
| Volume/Å$^3$ | 410.99(12) | 410.30(9) | 408.32(4) | 406.08(7) | 404.85(9) |
| Final R indexes [I>=2σ (I)] | $R_1$ = 2.01% | $R_1$ = 1.61% | $R_1$ = 1.51% | $R_1$ = 1.47% | $R_1$ = 1.27% |

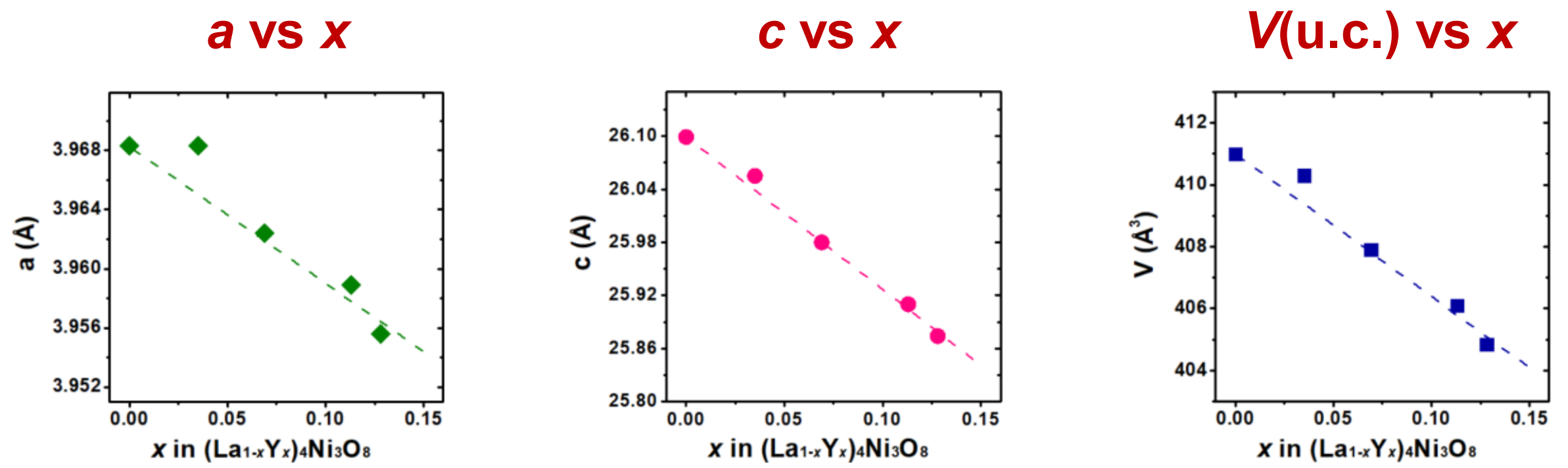

## II. Systematic reduction and oxidation experiments on $(La_{1-x}Pr_x)_4Ni_3O_{10}$ and $(La_{1-x}Pr_x)_4Ni_3O_8$ ($x$ = 0.55)

We conducted two control experiments. In the first, $(La_{1-x}Pr_x)_4Ni_3O_{10}$ (LPNO-4310) was gradually reduced to $(La_{1-x}Pr_x)_4Ni_3O_8$ (LPNO-438) ($x$=0.55) by controlling the time spent in the reducing environment. In the second, fully reduced LPNO-438 was marginally re-oxidized and subsequently reduced again. Each experimental step and the overall process are described in Fig. S1. Changes induced by the reduction and oxidation processes were monitored through AC susceptibility measurements with a 1 Oe magnitude field oscillating at 1 kHz.

In the first experiment, after AC susceptibility measurements were performed on a pristine LPNO-4310 single crystal, it was placed in a tube furnace under flowing Ar/3.5% $H_2$ gas with optimized flow rate. The furnace was heated up to 340 °C at a ramp rate of 0.5 °C/min, held at this temperature for 12 hours, and cooled at 0.5 °C/min. The heating and cooling rates were kept identical for all reduction steps, while the dwell time was systematically varied. After each reduction step, the AC susceptibility was measured again to track the evolution of the magnetic response. A total of three successive reduction steps were performed, with cumulative reduction times reaching 48 hours after the third step. Further reduction after 48 hours barely affected the AC susceptibility of the sample except for slightly increasing the overall magnetic moment due to increased ferromagnetic Ni particles appearing as a byproduct of the reduction process. The corresponding series of AC susceptibility measurements is shown in Fig. S2.

In Fig. S2(a), the AC susceptibility of pristine LPNO-4310 is presented. It shows a gradual Curie-like increase of the susceptibility with decreasing temperature, similar to that of pure $Pr_4Ni_3O_{10}$, indicating that the magnetic response in this system is dominated by the $Pr^{3+}$ cation. After the first reduction (Fig. S2(b)), the magnitude of the overall susceptibility has slightly decreased, while a hump emerges at ~ 50 K. After the second reduction (Fig. S2(c)), the hump becomes significantly more pronounced, which is presumably due to the formation of charge/spin-stripe ordering in LPNO-438. Further reduction strengthens the magnetic response of the sample, which is indicative of increased ferromagnetic Ni nanoparticle precipitates. Notably, the superconductivity emerges after the third step, which is 48 hours of reduction, as presented in Fig. S2(d).

In the second experiment, AC susceptibility measurements were conducted on fully reduced LPNO-438 placed in the tube furnace under flowing 100 sccm $O_2$ gas. The furnace was heated up to 200 °C at a ramp rate of 0.5 °C/min, held at this temperature for 2 hours, and cooled at 0.5 °C/min. The dwell time, and heating and cooling rates, were kept the same for all oxidation steps, while the dwell temperatures were varied. Similar to the reduction experiments, after each oxidation step, AC susceptibility measurements were conducted to monitor the change of the magnetic response. A total of three successive oxidation steps with temperatures 200°C, 300°C, 350°C, and the last reduction steps with the same protocols done for the reduction experiments with 48 hours were performed. The corresponding series of AC susceptibility measurements is shown in Fig. S3.

In Fig. S3, fully reduced LPNO-438 shows a clear diamagnetic response below 7 K, indicating the onset of superconductivity in this crystal. After the first oxidation step (200 °C), the diamagnetic response is nearly unchanged. In the second oxidation step with an elevated temperature (300 °C), the diamagnetic response has been significantly suppressed, and it nearly vanished after the third oxidation step (350 °C). While the magnetic background hasn't been significantly affected by these marginal oxidation steps, the sudden suppression of superconductivity unambiguously shows that the superconductivity of LPNO-438 and LYNO-438 is highly sensitive to the oxygen content in these systems.

Reduction steps

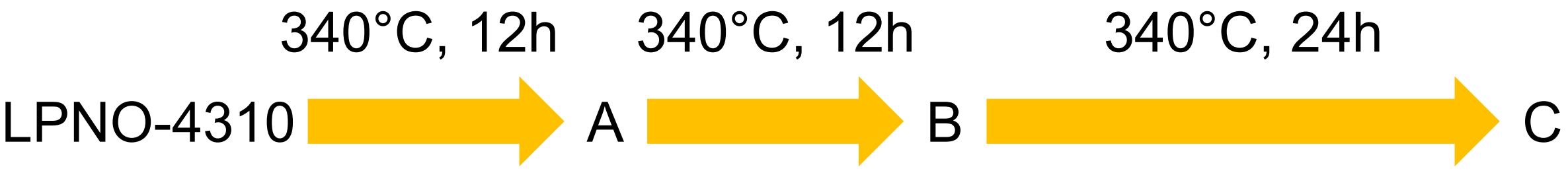

Oxidation steps

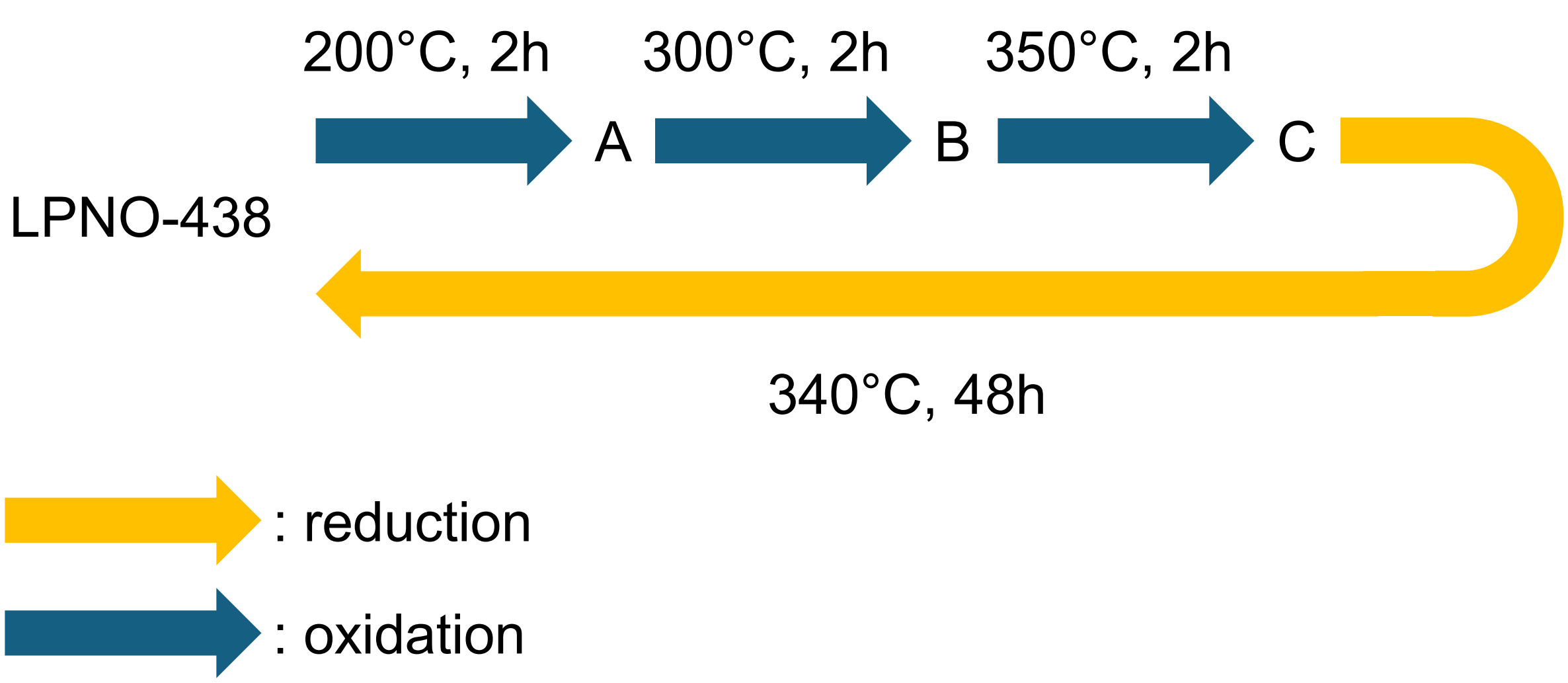

Fig. S1: Protocols for the reduction and oxidation experiments.

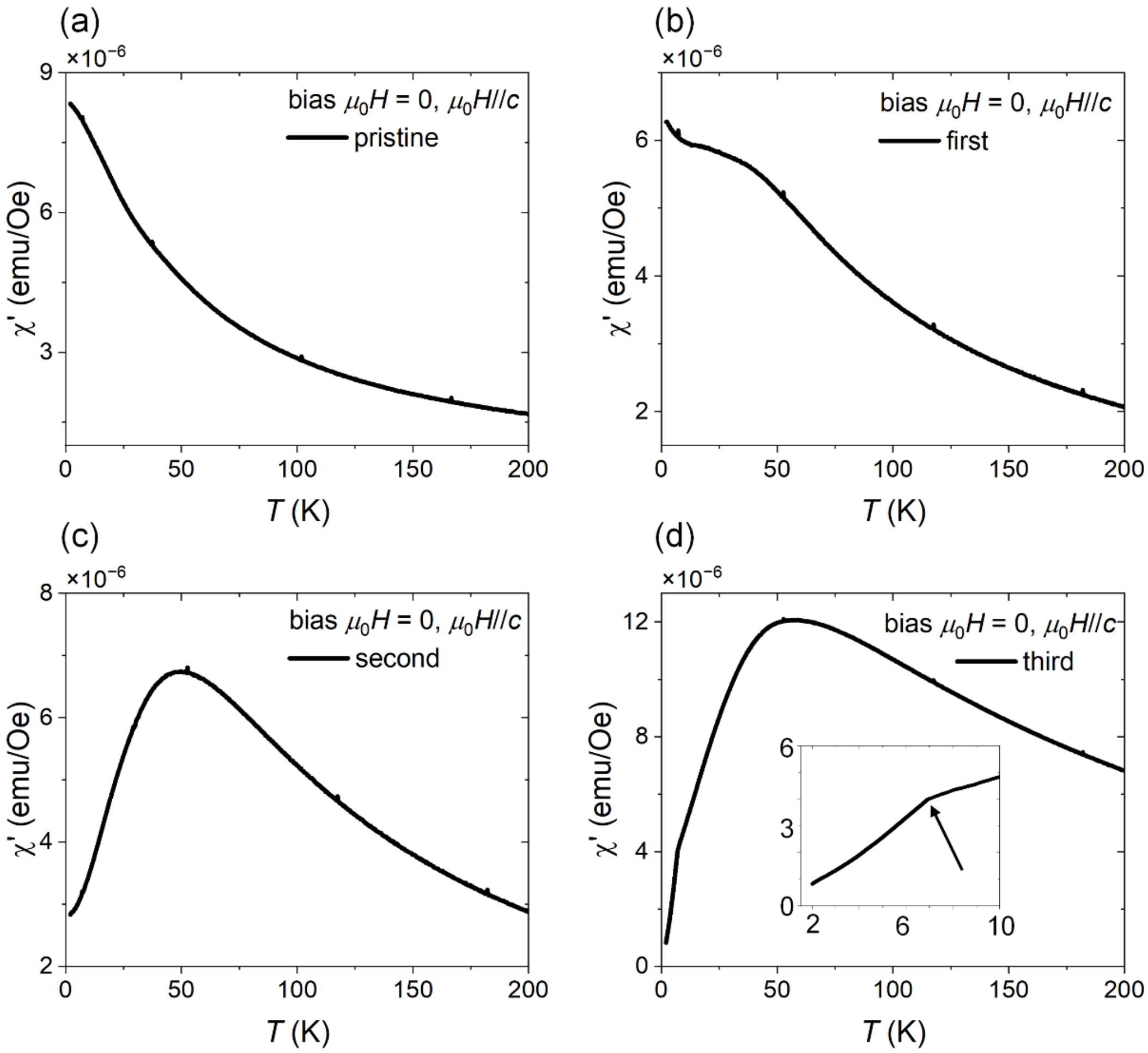

Fig. S2: AC susceptibility measurements of (a) pristine LPNO-4310 ($x$=0.55), (b) first-reduced, (c) second-reduced, and (d) third-reduced LPNO-438 ($x$=0.55) with zero DC bias field. The inset in (d) clearly shows the emerging diamagnetic response at ~ 7 K.

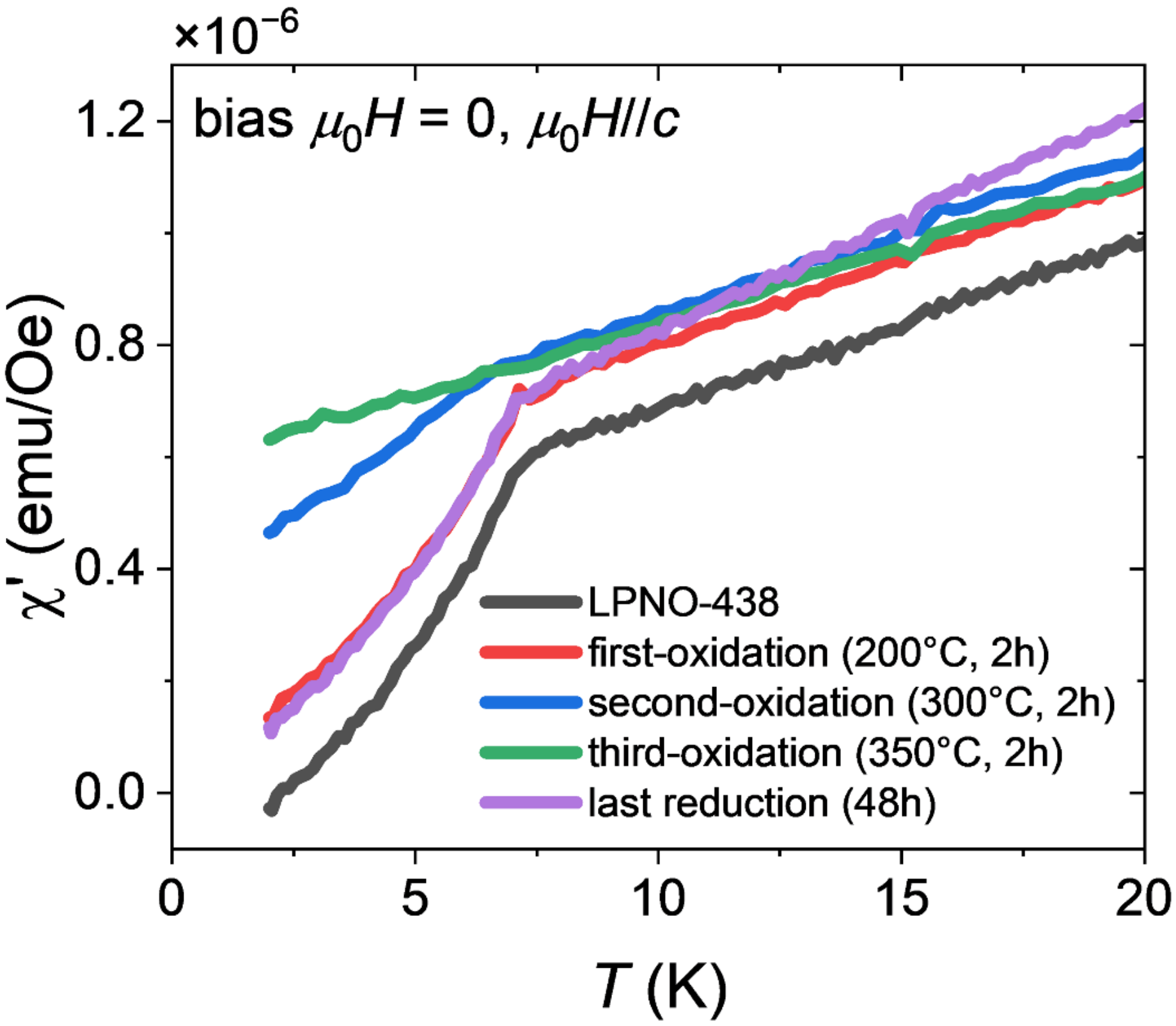

Fig. S3: AC susceptibility measurements of LPNO ($x$=0.55) with successive oxidation experiments and the last reduction step with a zero DC bias field.